\documentclass[aps,prb,reprint,superscriptaddress]{revtex4-2}
\usepackage[english]{babel}                     % English
\usepackage[T1]{fontenc}                        % naprednejše kodiranje fonta
\usepackage[utf8]{inputenc}                     % pravilno razpoznavanje unicode znakov
\usepackage{amsmath,amssymb,amsfonts,amsthm}    % matematični paketi
\usepackage{commath}
\usepackage{url}                                % \url and \href for links
\usepackage{bm}                                 % bold math
\usepackage{graphicx}
\usepackage{hyperref}
\hypersetup{
    colorlinks=true,   % false: boxed links; true: colored links
    linkcolor=blue,    % color of internal links
    citecolor=blue, % color of links to bibliography
    filecolor=magenta, % color of file links
    urlcolor=magenta,     % color of external links
    runcolor=cyan
}

\usepackage{times}
\usepackage[autostyle]{csquotes}
\usepackage{xcolor}
\allowdisplaybreaks % page break in Equations

\usepackage{siunitx}

\usepackage{cleveref}
\crefname{appendix}{Appendix}{Appendices}
\crefname{equation}{Eq.}{Eqs.}
\crefname{figure}{Fig.}{Figs.}
\crefname{table}{Table}{Tables}
\crefname{section}{Sec.}{Secs.}
\Crefname{equation}{Equation}{Equations} % for the references in the beginning of sentences
\Crefname{figure}{Figure}{Figures}
\Crefname{section}{Section}{Sections}

\renewcommand{\paragraph}[1]{\vspace{0.2cm}{\textit{#1}---\!}} % state-of-the-art PRL style (S.S. Lee)

\newcommand{\iu}{\mathrm{i}}                            % imaginary unit
\renewcommand{\vec}[1]{{\boldsymbol{#1}}}
\DeclareMathOperator{\sign}{sign}

\makeatletter
\newsavebox{\@brx}
\newcommand{\llangle}[1][]{\savebox{\@brx}{\(\m@th{#1\langle}\)}%
  \mathopen{\copy\@brx\kern-0.5\wd\@brx\usebox{\@brx}}}
\newcommand{\rrangle}[1][]{\savebox{\@brx}{\(\m@th{#1\rangle}\)}%
  \mathclose{\copy\@brx\kern-0.5\wd\@brx\usebox{\@brx}}}
\makeatother

\usepackage{changes}

\begin{document}

%\preprint{}

\title{Linear response functions from inhomogeneous dynamical mean-field theory}


\author{Don Rolih}
%\thanks{}
% \altaffiliation{}
\affiliation{Jo\v{z}ef Stefan Institute, Jamova 39, SI-1000 Ljubljana, Slovenia}
\affiliation{Faculty of Mathematics and Physics, University of Ljubljana, Jadranska 19, SI-1000 Ljubljana, Slovenia}
%\affiliation{Center for Computational Quantum Physics, Flatiron Institute, 162 Fifth Avenue, New York, NY 10010, USA}

\author{Fabian B. Kugler}
\affiliation{Institute for Theoretical Physics, University of Cologne, Z\"ulpicher Str. 77, 50937 Cologne, Germany}


\date{\today}

\begin{abstract}
We present a method for calculating static lattice susceptibilities with inhomogeneous dynamical mean-field theory.
The method utilizes the response of the system to an external field and thereby circumvents two-particle vertex functions.
We demonstrate the success of our approach for the magnetic susceptibility of the square-lattice Hubbard model using the numerical renormalization group as an impurity solver and show that it compares well with the results obtained using the standard vertex approach.
As it avoids vertex functions and is compatible with virtually any impurity solver, our method is able to reach low temperatures that are hard to access with other methods.
\end{abstract}

\maketitle

\section{Introduction}
\label{sec:introduction}
Strongly correlated electron systems exhibit a rich variety of phenomena, such as magnetism, superconductivity, and metal-insulator transitions.
To understand these phenomena,
two-particle correlation functions, such as susceptibilities, are a crucial tool: They describe the response of the system to external perturbations and characterize its collective excitations.
Dynamical mean-field theory (DMFT) allows one to study electronic correlations nonperturbatively~\cite{georges1996_dmft_RevModPhys}.
Within DMFT, the standard approach to computing lattice susceptibilities involves the local irreducible vertex function obtained from the impurity problem and solving the lattice Bethe--Salpeter equation (BSE).
Recent variations include the dual BSE~\cite{loon2024_dual_bethe-salpeter_equation_multiorbital_lattice_susceptibility_DMFT_PhysRevB} and methods for computing the static response at specific momenta~\cite{Strand2019}.
Nevertheless, the calculation of vertex functions to obtain generic lattice susceptibilities remains a computational hurdle, especially at low temperatures.

In another strand of development, extensions of DMFT to inhomogeneous systems have been used over the last two decades~\cite{potthoff1999a_surface_metal_transition_Hubbard_PhysRevB}. They have been successfully applied to a variety of problems, such as interfaces~\cite{potthoff1999c_metallic_surface_Mott_PhysRevB,potthoff2002_metal_insulator_transition_surfaces_InBook,helmes2008_kondo_proximity_effect_metal_into_mott_insulator_PhysRevLett}, heterostructures~\cite{freericks2004_DMFT_inhomogeneous_multilayered_PhysRevB,freericks2001_tuning_josephson_junction_quantum_critical_point_PhysRevB,zenia2009_appearance_fragile_fermi_liquids_finite_width_mott_insulators_idmft_PhysRevLett}, systems with impurities~\cite{chatterjee2019_real_space_DMFT_friedel_oscillations_PhysRevB}, and spin density waves in the two-dimensional (2D) Hubbard model~\cite{peters2014_spin_density_wave_hubbard_model_PhysRevB}.

Connecting these two developments, we present a method for calculating static susceptibilities with inhomogeneous DMFT (iDMFT). The method extracts the susceptibility from the linear response to a magnetic field, requiring only one-particle quantities from the impurity solver. We use an expansion in the applied field to avoid large matrix inversions and a resummation scheme that recovers magnetic instabilities in the thermodynamic limit. We also introduce a rank-1 approximation to the local vertex, which is an inexpensive improvement over the random phase approximation (RPA).
We show that the iDMFT method compares well with the results from the standard vertex approach on the paradigmatic example of the 2D Hubbard model and allows us to reach unprecedented, low temperatures using the numerical renormalization group (NRG) \cite{wilson1975_RG_Kondo_RevModPhys,bulla2008_nrg_RevModPhys} as the impurity solver.

The paper is organized as follows. In \cref{sec:model}, we introduce the model and formalism, including the standard vertex approach to computing lattice susceptibilities (\cref{sec:vertex_functions}), the applied-field method with iDMFT (\cref{sec:applied_field_dmft}), an expansion in the applied field that avoids numerically costly matrix inversions (\cref{sec:perturbative_expansion}), a resummation of the iDMFT equations that improves results in the thermodynamic limit (\cref{sec:resummation}), and we show the equivalence of the standard DMFT vertex and iDMFT approaches (\cref{sec:equivalence}). In \cref{sec:benchmark}, we benchmark our method against the standard vertex approach and obtain results at unprecedentedly low temperatures. We conclude in \cref{sec:conclusions}. We include several appendices: Appendix~\ref{sec:numerical_details} contains numerical details of the calculations, Appendix~\ref{sec:app_doping} further benchmarks to established results, Appendix~\ref{sec:app_rank1} a rank-1 approximation of the impurity vertex, and Appendix~\ref{sec:2d_dos} a derivation of the lattice density of states used in the main text.

\section{Formalism}
\label{sec:model}

\subsection{Hubbard model}
\label{sec:hubbard_model}
We consider the Hubbard model with Hamiltonian
\begin{equation}
H = \sum_{ij\sigma} t_{ij} c_{i\sigma}^\dagger c_{j\sigma}
+ \sum_{i\sigma} (V^{(i)}_\sigma - \mu)n_{i\sigma}
+ U \sum_i n_{i\uparrow} n_{i\downarrow}.
\end{equation}
Here, $c_{i\sigma}^\dagger$ creates an electron with spin $\sigma \in \{\uparrow,\downarrow\} \equiv \{ \pm 1 \}$ on site $i$, and $n_{i\sigma} = c_{i\sigma}^\dagger c_{i\sigma}$.
The chemical potential $\mu$ sets the average electron density,
$t_{ij}$ are the hopping amplitudes, and $U$ is the interaction strength. 
The single-particle potential $V^{(i)}_\sigma$ contains a local magnetic field $B$,
$V^{(i)}_\sigma = - \sigma B \delta_{i, 0}$, and thus breaks translational invariance.

We will consider both nearest-neighbor hopping $t$ and next-nearest-neighbor hopping $t'$ on the square lattice. The dispersion relation in equilibrium ($B=0$) is 
\begin{equation}
\label{eq:2d_dispersion}
\varepsilon_{\vec{k}} = - 2 t (\cos k_x + \cos k_y) - 4 t' \cos(k_x)\cos(k_y)
.
\end{equation}
The lattice density of states (with half-bandwidth $4t$ for $|t'|<t/2$) is~\cite{zhuravlev2011}
\begin{equation}
    \label{eq:2d_dos}
    \rho(\epsilon) =\frac{1}{2\pi^2 \sqrt{t^2 - t' \epsilon}}
    K \left(
    \sqrt{\frac{t^2 - (\epsilon + 4t')^2/16}{t^2 - \epsilon t'}}
    \right),
\end{equation}
where $K$ is the complete elliptic integral of the first kind. This simple formula holds only for $|t'| < t/2$; in Appendix~\ref{sec:2d_dos}, we generalize it to $|t'| > t/2$.

\subsection{Static susceptibility: vertex functions}
\label{sec:vertex_functions}

We aim to calculate the static lattice magnetic susceptibility
at inverse temperature $\beta = 1/T$,
\begin{subequations}
\begin{align}
\chi_i 
& =
\frac{\dif\, \langle S^z_i \rangle}{\dif B}\bigg|_{B=0}
\label{eq:physical_susceptibility_real_space}
=
2 \int_0^{\beta} \dif\tau\langle S_i^{z}(0) S_0^{z}(\tau)\rangle_{B=0}
\\
& = 
\int_0^{\beta} \dif\tau\langle n_{i\uparrow}(0) n_{0\uparrow}(\tau)
-
n_{i\uparrow}(0) n_{0\downarrow}(\tau)
\rangle_{B=0}
,
\end{align}
\end{subequations}
having used $S_{i}^{z} \!=\! \frac{1}{2}(n_{i\uparrow} - n_{i\downarrow})$
and SU(2) spin symmetry at $B \!=\! 0$.
The momentum-dependent counterpart follows from
\begin{equation}
\label{eq:physical_susceptibility_momentum_space}
\chi_{\vec{q}} = \sum_i e^{-\iu \vec{q} \cdot \vec{r}_i} \chi_i
.
\end{equation}
In single-site DMFT, the standard procedure for momentum-dependent susceptibilities is to compute the two-particle irreducible vertex $\Gamma_{\mathrm{imp}}$ of the impurity model and then solve the lattice BSE. Calculating $\Gamma_{\mathrm{imp}}$ is numerically costly, though, as it involves computing and inverting a two-particle Green function of the impurity model.

Let us start by establishing the notation. First, the one-particle Green functions on the impurity and on the lattice are%
\begin{subequations}
\begin{align}
G(\iu\nu) 
& = 
[1/\mathcal{G}_{0}(\iu\nu) -\Sigma(\iu\nu)]^{-1}
,
\\
G_{\vec{k}}(\iu\nu)
& =
[\iu\nu +\mu -\varepsilon_{\vec{k}}-\Sigma(\iu\nu)]^{-1}
,
\end{align}
\end{subequations}
respectively.
Here, $\mathcal{G}_{0}(\iu\nu)$ is the bare propagator (the ``Weiss field'') on the impurity and $\Sigma(\iu\nu)$ the self-energy.
With $G(\iu\nu)$ and $G_{\vec{k}}(\iu\nu)$, one can build the $\omega \!=\! 0$ two-particle propagators%
\begin{subequations}
\begin{align}
\chi_{0}^{\nu\nu'} 
& = -\beta\delta_{\nu\nu'}G(\iu\nu)^2
,
\\
\chi_{0, \vec{q}}^{\nu\nu'} 
& = 
- \beta \delta_{\nu\nu'} \frac{1}{N} \sum_{\vec{k}} G_{\vec{k} + \vec{q}}(\iu\nu) G_{\vec{k}}(\iu\nu)
,
\end{align}
\end{subequations}
on the impurity and on the lattice, respectively.
These two-frequency objects are \emph{elementary} generalized susceptibilities. Summing over the two frequencies subsequently, we obtain%
\begin{subequations}
\begin{align}
\chi_{0}^{\nu}
& = 
\frac{1}{\beta} \sum_{\nu'}
\chi_{0}^{\nu\nu'} 
= - G(\iu\nu)^2
, 
\\
\chi_{0}
& =
\frac{1}{\beta} \sum_{\nu}
\chi_{0}^{\nu}
=
- 
\frac{1}{\beta} \sum_{\nu}
G(\iu\nu)^2
,
\end{align}
\end{subequations}
where $\chi_{0}$ is the physical static susceptibility
on the impurity at $U \!=\! 0$ (implying $G \!=\! \mathcal{G}_0$). The analogous lattice objects are%
\begin{subequations}
\label{eq:lattice_elementary_susceptibilities}
\begin{align}
\chi_{0, \vec{q}}^\nu 
& =
\frac{1}{\beta} \sum_{\nu'}
\chi_{0, \vec{q}}^{\nu\nu'} 
= 
- \frac{1}{N} \sum_{\vec{k}} G_{\vec{k} + \vec{q}}(\iu\nu) G_{\vec{k}}(\iu\nu)
,
\\
\chi_{0, \vec{q}} 
& = 
\frac{1}{\beta} \sum_\nu
\chi_{0, \vec{q}}^\nu 
=
- \frac{1}{\beta N} \sum_{\nu \vec{k}} G_{\vec{k} + \vec{q}}(\iu\nu) G_{\vec{k}}(\iu\nu)
.
\end{align}
\end{subequations}

The \emph{full} static susceptibility has a similar structure. First,
\begin{equation}
\chi
=
\frac{1}{\beta} \sum_{\nu}
\chi^{\nu}
,
\qquad
\chi^{\nu}
= 
\frac{1}{\beta} \sum_{\nu'}
\chi^{\nu\nu'} 
\end{equation} 
on the impurity.
The (full) generalized magnetic susceptibility at $\omega \!=\! 0$, $\chi^{\nu\nu'}$, can be computed by an impurity solver from a slice of the two-particle Green function $G^{(2,\mathrm{ph})}_{\sigma\sigma'}(\iu\nu,\iu\nu',\iu\omega=0)$, parametrized in the particle-hole convention,
\begin{equation}
\label{eq:generalized_chi_ph}
\chi^{\nu\nu'} = G^{(2, \mathrm{ph})}_{\uparrow\uparrow}(\iu\nu, \iu\nu', \iu\omega=0) - G^{(2, \mathrm{ph})}_{\uparrow\downarrow}(\iu\nu, \iu\nu', \iu\omega=0)
.
\end{equation} 
Now, $\chi^{\nu\nu'}$ fulfills the impurity BSE~\cite{triqs_tprf}
\begin{equation}
\label{eq:bethe_salpeter_impurity}
\chi^{\nu\nu'} =
\chi_{0}^{\nu\nu'} 
- \frac{1}{\beta^2} \sum_{\nu_1, \nu_2} \chi_{0}^{\nu\nu_1} \Gamma_{\mathrm{imp}}^{\nu_1\nu_2} \chi^{\nu_2\nu'}
,
\end{equation}
where 
$\Gamma_{\mathrm{imp}}^{\nu_1\nu_2} = \Gamma_{\mathrm{imp}\uparrow\uparrow}^{\nu_1\nu_2} - \Gamma_{\mathrm{imp}\uparrow\downarrow}^{\nu_1\nu_2}$ is the two-particle irreducible vertex in the magnetic channel at vanishing transfer frequency.
It can be deduced from Eq.~\eqref{eq:bethe_salpeter_impurity} by matrix inversion,
$\vec{\Gamma}_{\mathrm{imp}} = \vec{\chi}^{-1} - \vec{\chi}_0^{-1}$.
The same local irreducible vertex is then used for the lattice BSE at $\omega=0$,
\begin{equation}
\label{eq:bethe_salpeter_lattice}
\chi_{\vec{q}}^{\nu\nu'} =
\chi_{0, \vec{q}}^{\nu\nu'} 
- \frac{1}{\beta^2} \sum_{\nu_1, \nu_2} \chi_{0, \vec{q}}^{\nu\nu_1} \Gamma_{\mathrm{imp}}^{\nu_1\nu_2} \chi_{\vec{q}}^{\nu_2\nu'}
.
\end{equation}
Finally, the physical susceptibility on the lattice follows from
\begin{equation}
\label{eq:physical_susceptibility_vertex}
\chi_{\vec{q}}
=
\frac{1}{\beta} \sum_{\nu}
\chi_{\vec{q}}^{\nu}
,
\qquad
\chi_{\vec{q}}^{\nu}
= 
\frac{1}{\beta} \sum_{\nu'}
\chi_{\vec{q}}^{\nu\nu'} 
.
\end{equation} 
One of the sums can already be done in Eq.~\eqref{eq:bethe_salpeter_lattice}, so that
\begin{equation}
\label{eq:bethe_salpeter_lattice_summed}
\chi_{\vec{q}}^{\nu} =
\chi_{0, \vec{q}}^{\nu} 
- 
\chi_{0,\vec{q}}^{\nu}
\frac{1}{\beta} \sum_{\nu'}  \Gamma_{\mathrm{imp}}^{\nu\nu'} \chi_{\vec{q}}^{\nu'}
.
\end{equation}

\subsection{Static susceptibility: external field in iDMFT}
\label{sec:applied_field_dmft}

The static susceptibility can also be obtained from the response to a small but finite magnetic field, localized at %site 
$i=0$:
\begin{equation}
\label{eq:susceptibility}
\chi_i
=
\frac{\dif\, \langle S^z_i \rangle}{\dif B}\bigg|_{B=0}
=
\lim_{B \to 0}
\frac{\langle S^z_i \rangle}{B}.
\end{equation}
We can use iDMFT to calculate $\langle S^z_i \rangle$ on each lattice site in the presence of $V^{(i)}_\sigma = - \sigma B \delta_{i, 0}$. The lattice Green function in real space, written in boldface to indicate the matrix structure and as a function of a complex frequency $z$, is given by
\begin{equation}
\label{eq:gfinverse_interacting}
\big[\mathbf{G}_{\sigma}^{-1}(z)\big]_{ij} = (z + \mu - V^{(i)}_\sigma - \Sigma^{(i)}_{\sigma}(z) )\delta_{ij} - t_{ij}
.
\end{equation}
The self-energy is local but site-dependent, i.e., $[\mathbf{\Sigma}_{\sigma}(z)]_{ij} = \Sigma_\sigma^{(i)}(z)\delta_{ij}$. 
It is computed from a set of impurity models, one for each (nonequivalent) lattice site.
Each impurity model is defined by a bare propagator (the ``Weiss field'') $\mathcal{G}_{0\sigma}^{(i)}(z)$.
It is determined self-consistently by the standard DMFT condition imposed on each lattice site,
\begin{equation}
\label{eq:real-space_dmft_update}
\frac{1}{\mathcal{G}_{0\sigma}^{(i)}(z)} =
\frac{1}{\big[\mathbf{G}_{\sigma}^{-1}(z)\big]_{ii}}
+ \Sigma_\sigma^{(i)}(z)
.
\end{equation}

The two numerically challenging steps in iDMFT are (i) solving each (nonequivalent) impurity model and (ii) obtaining the lattice Green function by inverting the matrix $\left[\mathbf{G}_{\sigma}^{-1}(z)\right]$ in Eq.~\eqref{eq:gfinverse_interacting}. Next, we show how matrix inversion can be avoided using an expansion in $B$.

\subsection{Expanding iDMFT in external field}
\label{sec:perturbative_expansion}

Without external field, the lattice Green function in single-site DMFT is spin-independent and diagonal in momentum,
\begin{equation}
\label{eq:interacting_lattice_green_k_space}
G_{\vec{k}}(z) = \frac{1}{z + \mu - \Sigma(z) - \varepsilon_{\vec{k}}}
.
\end{equation}
Inverse Fourier transform yields its real-space counterpart
\begin{equation}
\label{eq:fourier_transform_G}
[\mathbf{G}(z)]_{ij} = G_{i-j}(z) =
\frac{1}{N}
\sum_{\vec{k}} e^{\iu \vec{k}\cdot (\vec{r}_i - \vec{r}_j)} G_{\vec{k}}(z).
\end{equation}
By translational invariance, the elements of $\mathbf{G}(z)$ depend only on the difference $i-j$ of their row and column indices as indicated by the notation $G_{i-j}(z)$. 

To find the linear response, we can expand all iDMFT equations to linear order in $B$.
This allows us to exploit the numerically robust single-site DMFT as much as possible and avoid matrix inversions that are numerically costly and prone to finite-size artifacts.
The $\mathit{O}(B)$ contributions to the interacting lattice Green function $\mathbf{G}_\sigma(z)$ come from the single-particle potential $[\mathbf{V}_\sigma]_{ij} = V^{(i)}_\sigma \delta_{ij}$ and the self-energy.
The latter can be decomposed into the single-site DMFT contribution $\Sigma(z)$ and a spin- and site-dependent correction $\tilde\Sigma_\sigma^{(i)}(z) \sim \mathit{O}(B)$,
\begin{equation}
\label{eq:site_dependent_self_energy}
\Sigma_\sigma^{(i)}(z) = \Sigma(z) + \tilde\Sigma_\sigma^{(i)}(z)
.
\end{equation}
Thus, $\mathbf{G}_\sigma(z)$ can be written as
\begin{align}
%\label{eq:interacting_lattice_green_real_space_external_field}
\mathbf{G}_{\sigma}(z) &= [\mathbf{G}_{0}^{-1}(z) - \mathbf{V}_{\sigma} - \mathbf{\Sigma}_{\sigma}(z)]^{-1}
\nonumber
\\
%&= [\mathbf{G}_{0}^{-1}(\omega) - \mathbf{h}_{\sigma} - \tilde{\mathbf{\Sigma}}_{\sigma}(\omega) - \mathbf{\Sigma}(\omega)]^{-1}\\
&= [\mathbf{G}^{-1}(z)- \mathbf{V}_{\sigma} - \tilde{\mathbf{\Sigma}}_{\sigma}(z)]^{-1}
\nonumber
\\
\label{eq:perturbative_expansion_G_int}
& = \mathbf{G}(z) + \mathbf{G}(z) [\mathbf{V}_{\sigma} + \tilde{\mathbf{\Sigma}}_{\sigma}(z)] \mathbf{G}(z) + \mathit{O}(B^2)
.
\end{align}
The main advantage of Eq.~\eqref{eq:perturbative_expansion_G_int} is that we avoid inversion of the large matrix $\mathbf{G}^{-1}_\sigma(z)$, generally~\footnote{Using a selected inversion algorithm~\cite{Lin2011}, the diagonal elements $[\mathbf{G}_{\sigma}(z)]_{ii}$ can actually be obtained more efficiently.} \nocite{Lin2011} scaling as $\mathit{O}(N^3)$. The diagonal elements of Eq.~\eqref{eq:perturbative_expansion_G_int} are given by
\begin{align}
\label{eq:diagonal_elements_G_int}
[\mathbf{G}_{\sigma}(z)]_{ii} = G_{\mathrm{loc}}(z)
\!+\! \sum_{j} G_{i-j}(z) [V^{(j)}_\sigma \!+\! \tilde\Sigma^{(j)}_\sigma(z)] G_{j-i}(z)
.
\end{align}
Here, $G_{\mathrm{loc}}(z) = [\mathbf{G}(z)]_{ii}$ is the single-site DMFT local Green function. 
The computational cost of Eq.~\eqref{eq:diagonal_elements_G_int} scales as $\mathit{O}(N^2)$.
Furthermore, finite-size effects introduced by truncating the sum over $j$ are mild because the leading contribution $G_{\mathrm{loc}}(z)$ is taken from the thermodynamic limit.

The iDMFT update can also be expanded to first order in $B$. Inserting Eqs.~\eqref{eq:diagonal_elements_G_int} and \eqref{eq:site_dependent_self_energy}
into Eq.~\eqref{eq:real-space_dmft_update},
we find
\begin{align}
\frac{1}{\mathcal{G}_{0\sigma}^{(i)}(z)}
& =
\frac{1}{G_{\mathrm{loc}}(z)}
- \sum_j G_{i-j}(z) \frac{V^{(j)}_\sigma+\tilde\Sigma^{(j)}_\sigma(z)}{G_{\mathrm{loc}}(z)^2} G_{j-i}(z)
\nonumber
\\
& \ 
+ \Sigma(z) + \tilde\Sigma_\sigma^{(i)}(z)
+ \mathit{O}(B^2)
.
\end{align}
We can identify $\mathcal{G}_0(z) = 1/G_{\mathrm{loc}}(z) + \Sigma(z)$ from single-site DMFT and insert $V^{(i)}_\sigma = - \sigma B \delta_{i, 0}$, so that
\begin{align}
\frac{1}{\mathcal{G}_{0\sigma}^{(i)}(z)}
& =
\frac{1}{\mathcal{G}_0(z)}
+ \sigma B
\frac{G_{i}(z)G_{-i}(z)}{G_{\mathrm{loc}}(z)^2}
\nonumber
\\
& \
- \sum_{j \neq i} \tilde\Sigma^{(j)}_\sigma(z) \frac{G_{i-j}(z)G_{j-i}(z)}{G_{\mathrm{loc}}(z)^2} 
.
\label{eq:iDMFT_Weiss_field}
\end{align}

The magnetic susceptibility can still be computed from simple expectation values as $\chi_{i} = \langle S^z_i \rangle/B$ at small $B \neq 0$. 
Alternatively, we can decompose $\chi_{i}$ into a bubble part and vertex corrections, where the former are readily computed in the thermodynamic limit.
To this end, we write 
($B \to 0$ implicit)
\begin{equation}
\chi_{i} = \frac{\langle S^z_i \rangle}{B} 
= \frac{1}{2B \beta} \sum_{\nu} e^{\iu\nu 0^+} ([\mathbf{G}^\uparrow (\iu\nu)]_{ii} - [\mathbf{G}^\downarrow (\iu\nu)]_{ii})
.
\end{equation}
Note that we can thus identify $\chi_i^\nu = ([\mathbf{G}^\uparrow (\iu\nu)]_{ii} - [\mathbf{G}^\downarrow (\iu\nu)]_{ii})/(2B)$.
We then insert Eq.~\eqref{eq:diagonal_elements_G_int}, $V^{(i)}_\sigma = - \sigma B \delta_{i, 0}$, and abbreviate $\delta\Sigma^{(i)} = \Sigma^{(i)}_\uparrow-\Sigma^{(i)}_\downarrow = \tilde\Sigma^{(i)}_\uparrow-\tilde\Sigma^{(i)}_\downarrow$ to get
\begin{equation}
\label{eq:chi_iDMFT_bubble_vertex}
\chi_i = -\frac{1}{\beta}\sum_{j\nu}
G_{i - j}(\iu\nu) 
\bigg[
\delta_{j,0} - \frac{\delta\Sigma^{(j)}(\iu\nu)}{2B}
\bigg]
G_{j-i}(\iu\nu)
.
\end{equation}
Next, we transform to momentum space via Eqs.~\eqref{eq:physical_susceptibility_momentum_space} and \eqref{eq:fourier_transform_G}:
\begin{align}
\chi_{\vec{q}} 
& = 
\frac{-1}{\beta N^2} \sum_{ij\nu\vec{k}\vec{k}'} e^{-\iu \vec{q} \cdot \vec{r}_i} e^{\iu (\vec{k}'-\vec{k}) \cdot (\vec{r}_i - \vec{r}_j)} G_{\vec{k}'}(\iu\nu) G_{\vec{k}}(\iu\nu)
\nonumber \\
& \hspace{3cm} \times
[
\delta_{j,0} - \delta\Sigma^{(j)}(\iu\nu)/(2B)]
.
\end{align}
The sum over $i$ yields $N \delta_{\vec{k}',\vec{k}+\vec{q}}$, so that
\begin{subequations}
\label{eq:chi_iDMFT_bubble_vertex_momentum_space}
\begin{align}
\chi_{\vec{q}} 
& = 
\chi_{0,\vec{q}}
-
\frac{1}{\beta}\sum_{\nu} \chi_{0,\vec{q}}^{\nu} \frac{\delta\Sigma_{\vec{q}}(\iu\nu)}{2B}
,
\\
\delta\Sigma_{\vec{q}}(\iu\nu)
& = \sum_i e^{-\iu \vec{q} \cdot \vec{r}_i} \delta\Sigma^{(i)}(\iu\nu)
.
\end{align}
\end{subequations}
where we used $\chi_{0,\vec{q}}$ and $\chi_{0,\vec{q}}^\nu $ from Eq.~\eqref{eq:lattice_elementary_susceptibilities}.
One sees that $\chi_{0,\vec{q}}$ provides the bubble contribution to $\chi_{\vec{q}}$,
while $\delta\Sigma_{\vec{q}}(\iu\nu)$ encodes the vertex corrections
(see also \cref{sec:equivalence}).

Computing the susceptibility from $\chi_{i} = \langle S^z_i \rangle/B$ or from Eq.~\eqref{eq:chi_iDMFT_bubble_vertex_momentum_space} yielded equivalent numerical results for all cases considered. Indeed, the vertex corrections typically dominate the magnetic susceptibility, so that the gain of obtaining the bubble part in the thermodynamic limit is mild. 
Nevertheless, the decomposition of Eq.~\eqref{eq:chi_iDMFT_bubble_vertex_momentum_space} is helpful for interpretation purposes, and it allows one to construct a resummation scheme (with important numerical consequences), as explained next.

\subsection{Resummation of the iDMFT expansion}
\label{sec:resummation}

In a finite system, $\chi_{\vec{q}}$ will always be finite.
This corresponds to the fact that, with a finite number of $\delta \Sigma^{(j)}(\iu\nu)$ terms, the vertex corrections encoded in $\delta\Sigma_{\vec{q}}(\iu\nu)$ will remain finite.
To describe a magnetic phase transition in the thermodynamic limit, a resummation (similar to the BSE in the vertex approach) is needed.

To understand the structure, let us first separate the Hartree contribution $U n_{-\sigma}$ from the self-energy $\Sigma^{(i)}_\sigma(\iu\nu)$.
Accordingly, we write $\delta\Sigma^{(i)}(\iu\nu) = -2U \langle S^z_i \rangle + \delta\hat\Sigma^{(i)}(\iu\nu)$, where the hat indicates the \enquote{non-Hartree} part.
We obtain $\delta\Sigma_{\vec{q}}(\iu\nu)
= 
- 2 B U \chi_{\vec{q}}
+ \delta\hat\Sigma_{\vec{q}}(\iu\nu)$ and thus%
\begin{subequations}
\label{eq:rpa_beyond}
\begin{align}
\chi_{\vec{q}} 
& = 
\chi_{0,\vec{q}} 
+ \chi_{0,\vec{q}} U \chi_{\vec{q}}
- \frac{1}{\beta}\sum_{\nu} \chi_{0,\vec{q}}^{\nu} \frac{\delta\hat\Sigma_{\vec{q}}(\iu\nu)}{2B} 
\\
\Leftrightarrow
\chi_{\vec{q}}
& = 
(1 - \chi_{0,\vec{q}} U)^{-1} \bigg[
\chi_{0,\vec{q}}
- \frac{1}{\beta}\sum_{\nu} \chi_{0,\vec{q}}^{\nu} \frac{\delta\hat\Sigma_{\vec{q}}(\iu\nu)}{2B} 
\bigg]
.
\label{eq:iDMFT_RPA_resummation}
\end{align}
\end{subequations}
We identify the first part as the RPA series and the second part as the vertex corrections beyond RPA.
Through the denominator, Eq.~\eqref{eq:iDMFT_RPA_resummation} clearly enables a divergent susceptibility. However, since we only related the Hartree part of $\Sigma$ to $\chi_{\vec{q}}$, the divergence criterion reduces to the RPA Stoner criterion.

To improve the divergence criterion, we make an ansatz $\delta\Sigma_{\vec{q}}(\iu\nu) = -2B\lambda_{\vec{q}}(\iu\nu)\chi_{\vec{q}}$, inspired by the above analysis.
The resulting expression for the susceptibility is%
\begin{subequations}
\label{eq:chi_q_psi_corrected}
\begin{align}
\chi_{\vec{q}}
& = 
\chi_{0,\vec{q}}
+
\frac{1}{\beta}\sum_{\nu} \chi_{0,\vec{q}}^{\nu}
\lambda_{\vec{q}}(\iu\nu) \chi_{\vec{q}}
\\
\Leftrightarrow
\chi_{\vec{q}}
& =
\frac{\chi_{0,\vec{q}}}{1 - U P_{\vec{q}}}
,
\quad
P_{\vec{q}}
=
\frac{1}{U \beta}\sum_{\nu} \chi_{0,\vec{q}}^{\nu}
\lambda_{\vec{q}}(\iu\nu) 
.
\end{align}
\end{subequations}
The new divergence criterion is $UP_{\vec{q}}=1$,
expressed via the \enquote{polarization} $P_{\vec{q}}$.
One readily observes that $P_{\vec{q}}$ is the only quantity with system-size dependence.
So, we may write more explicitly $P_{\vec{q}}
=
\lim_{N\to\infty} P^N_{\vec{q}}$
with%
\begin{subequations}
\begin{align}
P^N_{\vec{q}}
& =
\frac{-1}{2U \beta}\sum_{\nu} \chi_{0,\vec{q}}^{\nu}
\frac{\sum_{|i| \leq N} e^{-\iu \vec{q} \cdot \vec{r}_i} \delta\Sigma^{(i)}(\iu\nu)}{\sum_{|i| \leq N} e^{-\iu \vec{q} \cdot \vec{r}_i} \langle S^z_i \rangle}.
\\
& =
\chi_{0,\vec{q}}
-
\frac{1}{2U \beta}\sum_{\nu} \chi_{0,\vec{q}}^{\nu}
\frac{\sum_{|i| \leq N} e^{-\iu \vec{q} \cdot \vec{r}_i} \delta\hat\Sigma^{(i)}(\iu\nu)}{\sum_{|i| \leq N} e^{-\iu \vec{q} \cdot \vec{r}_i} \langle S^z_i \rangle}.
\end{align}
\end{subequations}
We will see that using $P_{\vec{q}}$ not only enables a divergent susceptibility, but improves the numerical results quite generally.

\subsection{Equivalence of vertex and external field approaches}
\label{sec:equivalence}

We want to show that the static susceptibilities computed from Eq.~\eqref{eq:bethe_salpeter_lattice_summed} and Eq.~\eqref{eq:chi_iDMFT_bubble_vertex_momentum_space} are equivalent. This is suspected as DMFT is $\Phi$-derivable and thus thermodynamically consistent~\cite{hafermann2014}. Indeed, $\Phi$-derivability will enter crucially through $\Gamma \propto \delta \Sigma / \delta G$.
First, we rewrite Eq.~\eqref{eq:chi_iDMFT_bubble_vertex_momentum_space} as ($B \to 0$ implicit)
\begin{align}
\label{eq:iDMFT_chi_q_nu}
\chi_{\vec{q}} 
= 
\frac{1}{\beta}\sum_{\nu}
\chi_{\vec{q}}^\nu
, \quad
\chi_{\vec{q}}^\nu 
= 
\chi_{0,\vec{q}}^{\nu}
\bigg[ 1 - 
\frac{\dif\Sigma_{\uparrow\vec{q}}(\iu\nu)}{\dif B}
\bigg]
,
\end{align}
and then express the self-energy derivative as
\begin{align}
\label{eq:Sigma_derivative}
\frac{\dif \Sigma_{\uparrow\vec{q}}(\iu\nu)}{\dif B}
& =
\frac{1}{N^2 \beta} \sum_{\sigma \nu' \vec{k} \vec{k}'}
\frac{\delta \Sigma_{\uparrow\vec{q}}(\iu\nu)}{\delta G_{\sigma \vec{k} \vec{k}'}(\iu\nu')}
\frac{\dif G_{\sigma \vec{k} \vec{k}'}(\iu\nu')}{\dif B}
.
\end{align}
For the first factor, we can use the functional Ward identity
\begin{align}
\label{eq:functional_Ward_id}
\frac{\delta \Sigma_{\uparrow\vec{q}}(\iu\nu)}{\delta G_{\sigma \vec{k} \vec{k}'}(\iu\nu')}
\bigg|_{B=0}
=
-
N\delta_{\vec{k}',\vec{k}+\vec{q}}
\Gamma_{\mathrm{imp}\uparrow\sigma}^{\nu\nu'}
.
\end{align}
The second term, using standard manipulations,  gives
\begin{align}
&
\frac{\dif G_{\sigma \vec{k} \vec{k}'}(\iu\nu')}{\dif B}
=
\frac{-1}{N^2} \sum_{\tilde{\vec{k}}\tilde{\vec{k}}'}
G_{\sigma \vec{k} \tilde{\vec{k}}}(\iu\nu')
\frac{\dif G^{-1}_{\sigma\tilde{\vec{k}}\tilde{\vec{k}}'}(\iu\nu)}{\dif B}
G_{\sigma \tilde{\vec{k}}' \vec{k}'}(\iu\nu')
\nonumber \\
& =
\frac{\sigma}{N^2} \sum_{\tilde{\vec{k}}\tilde{\vec{k}}'}
G_{\sigma \vec{k} \tilde{\vec{k}}}(\iu\nu')
\bigg[
\frac{\dif \Sigma_{\uparrow,\tilde{\vec{k}}'-\tilde{\vec{k}}}(\iu\nu')}{\dif B}
- 1
\bigg]
G_{\sigma \tilde{\vec{k}}' \vec{k}'}(\iu\nu')
\nonumber \\
& 
\xrightarrow[B \to 0]{}
\sigma
G_{\vec{k}}(\iu\nu')
\bigg[
\frac{\dif \Sigma_{\uparrow,\vec{k}'-\vec{k}}(\iu\nu')}{\dif B}
\bigg|_{B=0}
- 1
\bigg]
G_{\vec{k}'}(\iu\nu')
,
\label{eq:G_derivative}
\end{align}
where we used $\dif \Sigma_{\sigma\vec{q}}(\iu\nu') / \dif B = \sigma \dif \Sigma_{\uparrow\vec{q}}(\iu\nu') / \dif B$.
Inserting Eqs.~\eqref{eq:functional_Ward_id} and \eqref{eq:G_derivative} into Eq.~\eqref{eq:Sigma_derivative}, we obtain
\begin{align}
\frac{\dif \Sigma_{\uparrow\vec{q}}(\iu\nu)}{\dif B}
& =
\frac{1}{\beta} \sum_{\nu'}
\sum_\sigma \sigma
\Gamma_{\mathrm{imp}\uparrow\sigma}^{\nu\nu'}
\bigg[
\frac{\dif \Sigma_{\uparrow,\vec{q}}(\iu\nu')}{\dif B}
- 1
\bigg]
\nonumber \\
& \ \times
\frac{-1}{N} \sum_{\vec{k}}
G_{\vec{k}}(\iu\nu')
G_{\vec{k}+\vec{q}}(\iu\nu')
.
\end{align}
The spin sum produces
$\Gamma^{\nu\nu'}_{\mathrm{imp}} = \Gamma^{\nu\nu'}_{\mathrm{imp}\uparrow\uparrow} - \Gamma^{\nu\nu'}_{\mathrm{imp}\uparrow\downarrow}$,
the momentum sum $\chi_{0,\vec{q}}^\nu$, see Eq.~\eqref{eq:lattice_elementary_susceptibilities},
and with Eq.~\eqref{eq:iDMFT_chi_q_nu}, we get
\begin{align}
\frac{\dif \Sigma_{\uparrow\vec{q}}(\iu\nu)}{\dif B}
=
\frac{1}{\beta} \sum_{\nu'}
\Gamma_{\mathrm{imp}}^{\nu\nu'}
\chi_{\vec{q}}^{\nu'}
.
\label{eq:contracted_WI_1}
\end{align}
Inserting this result into Eq.~\eqref{eq:iDMFT_chi_q_nu}, we reproduce Eq.~\eqref{eq:bethe_salpeter_lattice_summed},
thus proving equivalence between both ways of computing $\chi_{\vec{q}}.$
In Appendix~\ref{sec:app_rank1}, we explore a simple rank-1 approximation of the impurity vertex based on \cref{eq:contracted_WI_1} and show that it improves upon RPA at the cost of a single extra run of the impurity solver in a magnetic field.

Finally, we may transform the Ward identity Eq.~\eqref{eq:contracted_WI_1}, involving a two-particle irreducible vertex and the full susceptibility, into an expression involving the full vertex and an elementary susceptibility. We first write Eq.~\eqref{eq:contracted_WI_1} as
\begin{align}
\frac{\dif \Sigma_{\uparrow\vec{q}}(\iu\nu)}{\dif B}
=
\frac{1}{\beta^2} \sum_{\nu'\nu''}
\Gamma_{\mathrm{imp}}^{\nu\nu'}
\chi_{\vec{q}}^{\nu'\nu''}
\label{eq:contracted_WI_2}
\end{align}
and manipulate the expression on the right in matrix notation,
\begin{align}
\vec{\Gamma}_{\mathrm{imp}} \vec{\chi}_{\vec{q}}
& =
\vec{\Gamma}_{\mathrm{imp}}
( \vec{\chi}_{0,\vec{q}} - \vec{\chi}_{0,\vec{q}} \vec{\Gamma}_{\mathrm{imp}} \vec{\chi}_{\vec{q}} )
,
\nonumber \\
\Leftrightarrow \
\vec{\Gamma}_{\mathrm{imp}} \vec{\chi}_{\vec{q}}
& =
(1 + \vec{\Gamma}_{\mathrm{imp}} \vec{\chi}_{0,\vec{q}} )^{-1}
\vec{\Gamma}_{\mathrm{imp}}
\vec{\chi}_{0,\vec{q}}
=
\vec{F}_{\vec{q}}
\vec{\chi}_{0,\vec{q}}
.
\end{align}
In the first step, we used the BSE for the generalized susceptibility.
In the last step, we introduced the full vertex $\vec{F} \!=\! \vec{F}_{\uparrow\uparrow} - \vec{F}_{\uparrow\downarrow}$ through its BSE
\begin{align}
\vec{F}_{\vec{q}}
= 
\vec{\Gamma}_{\mathrm{imp}}
\!-\!
\vec{\Gamma}_{\mathrm{imp}}
\vec{\chi}_{0,\vec{q}}
\vec{F}_{\vec{q}}
\,
\Leftrightarrow
\,
\vec{F}_{\vec{q}}
=
(1 \!+\! \vec{\Gamma}_{\mathrm{imp}} \vec{\chi}_{0,\vec{q}} )^{-1}
\vec{\Gamma}_{\mathrm{imp}}
.
\end{align}
We can thus rewrite the Ward identity, Eq.~\eqref{eq:contracted_WI_1}, as
\begin{align}
\frac{\dif \Sigma_{\uparrow\vec{q}}(\iu\nu)}{\dif B}
=
\frac{1}{\beta} \sum_{\nu'}
F_{\vec{q}}^{\nu\nu'}
\chi_{0,\vec{q}}^{\nu'}
.
\label{eq:contracted_WI_3}
\end{align}

\section{Numerical results}
\label{sec:benchmark}

To validate the iDMFT method, we benchmark it against established results for the single-band $t$-$t'$ square-lattice Hubbard model (Sec.~\ref{sec:hubbard_model}). 
To illustrate its advantages, we show results at unprecedented, low temperatures.
All results are given in units of $t=1$. We use NRG as the impurity solver for iDMFT (see Appendix~\ref{sec:numerical_details} for numerical details).

\subsection{Weak interaction at half-filling}
\label{sec:res_weak_coupling}

\begin{figure}
    \includegraphics[width=\columnwidth]{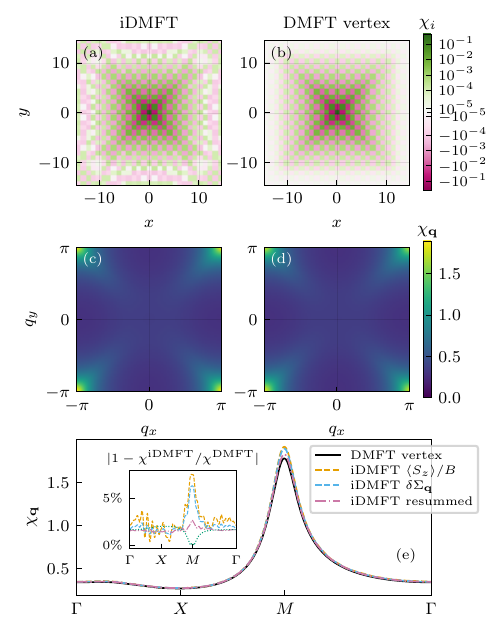}
    \caption{Lattice spin susceptibility of the half-filled 2D Hubbard model at $U=2$ and $\beta=5$ from (a, c) iDMFT  and (b, d) the DMFT vertex. Panels (a, b) show $\chi_i$ in real space and panels (c, d) $\chi_{\vec{q}}$ in momentum space. The iDMFT result in (a) is computed as $\langle S^z_i \rangle/B$ from the expectation value of the NRG impurity solver. Panel (e) shows $\chi_{\vec{q}}$ along a high-symmetry path obtained from the DMFT vertex, iDMFT via $\langle S^z_i\rangle/B$, iDMFT via $\delta \Sigma_{\vec{q}}$ [\cref{eq:chi_iDMFT_bubble_vertex_momentum_space}], and iDMFT resummation as defined in \cref{eq:chi_q_psi_corrected}. The inset shows the relative deviation of all three iDMFT results w.r.t.\ the DMFT vertex result.
    }
    \label{fig:chi_r_chi_path_U=2}
\end{figure}

We first consider half-filling, small $U$, and relatively high temperatures. Here, continuous-time quantum Monte Carlo (QMC) impurity solvers can accurately resolve the two-particle Green function $G^{(2)}_{\sigma\sigma'}$ required by the vertex approach, facilitating a direct comparison between the two methods. We use a continuous-time interaction-expansion (CT-INT) QMC solver~\cite{rubtsov2005,gull2011, triqs_ctint} to get $G^{(2)}_{\sigma\sigma'}$ and the Two-Particle Response Function Toolbox~\cite{triqs_tprf, parcollet2015} to solve the BSE.

\Cref{fig:chi_r_chi_path_U=2} %shows the lattice spin susceptibility 
presents our results
at $U=2$ and $\beta=5$. Panels (a, b) display the real-space susceptibility 
obtained from iDMFT and from the standard DMFT vertex calculation; panels (c, d) show the same data in momentum space.
Both methods reproduce the onset of an antiferromagnetic (AFM) instability with a checkerboard pattern  
in $\chi_i$ and a pronounced peak at the AFM ordering vector $(\pi, \pi)$ in $\chi_{\vec{q}}$.
The faint artifacts at the box boundaries in (a) are not a finite-size effect but stem from the fact that the magnitude of $\chi_i$ reaches the noise threshold of the impurity solver (see Appendix~\ref{sec:numerical_details} for further details).
Panel (e) compares $\chi_{\vec{q}}$ along the $\Gamma$-$X$-$M$-$\Gamma$ high-symmetry path.
All curves agree very well; the only quantitative difference is at $(\pi, \pi)$ where the iDMFT values are higher by a few percent. The inset in \cref{fig:chi_r_chi_path_U=2}(e) shows the relative difference between all three iDMFT results and the standard DMFT susceptibility. The resummed iDMFT of \cref{eq:chi_q_psi_corrected} decreases the difference between the two methods to about two percent.

\begin{figure}
    \includegraphics[width=\columnwidth]{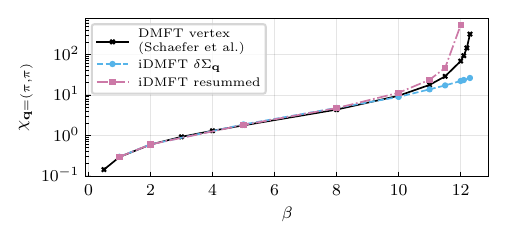}
    \caption{Lattice spin susceptibility at the AFM ordering vector $\vec{q}=(\pi, \pi)$ as a function of inverse temperature $\beta$ at half-filling and $U \!=\! 2$, computed with iDMFT via \cref{eq:chi_iDMFT_bubble_vertex_momentum_space} (blue) and iDMFT resummation via \cref{eq:chi_q_psi_corrected} (violet). Black points are DMFT results from Ref.~\onlinecite{schaefer2021} (using the lattice BSE and a CT-INT vertex).
    }
    \label{fig:chi_pipi_Schaefer}
\end{figure}

We examine the onset of AFM ordering in \Cref{fig:chi_pipi_Schaefer}, where we show $\chi_{\vec{q}}$ at the ordering vector $\vec{q} = (\pi, \pi)$ as a function of inverse temperature $\beta$ at $U=2$, along with the DMFT results of Ref.~\onlinecite{schaefer2021}. The latter were obtained from the lattice BSE with a CT-INT solver (we divided their data by a factor of two to match our convention). For $\beta \leq 8$, the iDMFT results match the reference data. 

Because DMFT is a mean-field theory, it predicts a finite-temperature AFM transition even in the 2D Hubbard model where long-range order is prohibited.
Hence, $\chi_{\vec{q}}$ obtained from the DMFT vertex diverges for $\beta \gtrsim 12$.
Increasing $\beta$ toward the AFM transition, the magnetic correlation length grows and eventually exceeds the system size chosen in our iDMFT approach. The iDMFT susceptibility of \cref{eq:chi_iDMFT_bubble_vertex_momentum_space} thus remains regular with no sign of a divergence.
This is remedied by the resummation of \cref{eq:chi_q_psi_corrected}, as demonstrated by the resummed iDMFT result shown in \cref{fig:chi_pipi_Schaefer} as violet points.

\subsection{Strong interaction at finite doping and $t'=0$}

\begin{figure}
    \centering
    \includegraphics[width=\columnwidth]{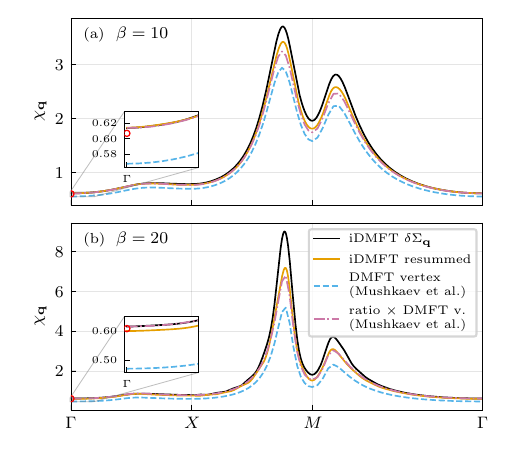}
    \caption{Lattice spin susceptibility at $U=8$, $t'=0$, $n=0.78$, and (a) $\beta=10$, (b) $\beta=20$. Results are calculated with iDMFT via \cref{eq:chi_iDMFT_bubble_vertex_momentum_space} (black) and \cref{eq:chi_q_psi_corrected} (orange) and compared to DMFT results from Ref.~\onlinecite{mushkaev2025}. We rescaled their raw data (blue dashed curve) with the ratio $\chi_{\vec{q}=0}^{\mathrm{DMFT}}/\chi_{\vec{q}=0}^{\mathrm{Mush.}}$, where $\chi_{\vec{q}=0}^{\mathrm{DMFT}}$ is the single-site DMFT result. The latter value is marked with a hollow red circle at the $\Gamma$ point and enlarged in the insets.
    }
    \label{fig:chi_q_path_idmft_vs_mushkaev}
\end{figure}

As a first test in a regime where QMC solvers typically struggle, we consider finite hole doping, $n=0.78$, at $U=8$. This parameter set was studied recently in Ref.~\onlinecite{mushkaev2025} in the context of the spin-stripe transition.

\Cref{fig:chi_q_path_idmft_vs_mushkaev} shows the lattice spin susceptibility along the high-symmetry path for two representative temperatures, $\beta=10,\,20$. The susceptibility peaks at an incommensurate wave vector $(\pi, \pi - 2 \pi \eta)$ with $\eta \approx 0.12$. The inset enlarges the behavior at $\vec{q}=0$, with the circle giving the (highly accurate) result from single-site DMFT solved with a finite $B$ field.

A direct comparison with $\chi_{\vec{q}}$ from Ref.~\onlinecite{mushkaev2025} in \cref{fig:chi_q_path_idmft_vs_mushkaev} reveals a quantitative discrepancy in magnitude, while the incommensurate peak position and the overall momentum dependence agree. We have confirmed with the authors of Ref.~\onlinecite{mushkaev2025} that this discrepancy is most likely due to the frequency truncation of the two-particle vertex in their calculation and that it does not affect their phase-diagram results. Rescaling their results to match the single-site DMFT value for $\chi_{\vec{q}=0}$ gives quantitative agreement with our resummed iDMFT results, see \cref{fig:chi_q_path_idmft_vs_mushkaev}.

\subsection{Strong interaction at finite doping and $t'=-0.2$}
\label{sec:rec_strong_coupling_tprime}

We now include
finite $t'=-0.2$ at $U=8$ and finite doping, $n=0.72$. At $\beta=12.5$, these parameters match those of Ref.~\onlinecite{vilardi2018}, where it was shown that the local but strongly frequency-dependent DMFT vertex shifts the dominant magnetic instability away from the 
AFM ordering vector $(\pi,\pi)$ favored by the particle-hole bubble toward incommensurate wave vectors of the form $(\pi, \pi-2\pi\eta)$. We first compare our results at $\beta=12.5$ to Ref.~\onlinecite{vilardi2018}. Then, we show that iDMFT allows us to obtain results at much lower temperatures too.

\begin{figure}
    \centering
    \includegraphics{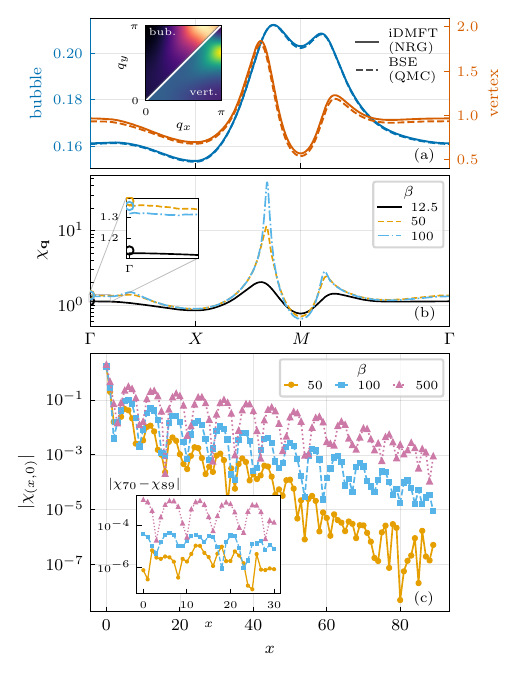}
    \caption{
    Lattice spin susceptibility at $U=8$, $t'=-0.2$, $n=0.72$ and various $\beta$.
    (a)~Bubble and vertex contributions to $\chi_{\vec{q}}$ along $\Gamma$-$X$-$M$-$\Gamma$ at $\beta \!=\! 12.5$ (i.e., the parameters of Ref.~\onlinecite{vilardi2018}). Inset: the same decomposition as a heat map 
    %over the Brillouin-zone quadrant split into two triangular regions
    in the irreducible Brillouin zone (bubble upper-left, vertex lower-right). The solid red curve for the vertex contribution is from resummed iDMFT with NRG as the impurity solver; the dashed red curve is the BSE result from a QMC impurity solver~\cite{wallerberger2019}.
    (b)~$\chi_{\vec q}$ along the high-symmetry path for $\beta=12.5, 50, 100$, reaching unprecedented, low temperatures. Circles mark the single-site DMFT susceptibility $\chi_{\vec{q}=0}$; the inset zooms in near $\Gamma$ to show that it coincides with the $\Gamma$-point value of the full lattice calculation. The curve for $\beta \!=\! 12.5$ is iDMFT resummed; curves for higher $\beta$ are iDMFT via \cref{eq:chi_iDMFT_bubble_vertex_momentum_space}.
    (c)~Real-space decay of $\chi_i$ for $\vec{r}_i = (x, 0)$ for a calculation with $d_{\mathrm{max}}=89$ (4095 independent impurities). The inset shows the difference between iDMFT calculations using $d_{\mathrm{max}}=70$ and $d_{\mathrm{max}}=89$ (2556 and 4095 independent impurities, respectively). Differently from panel (b), we here include results at $\beta=500$, too, which are not fully converged in $d_{\mathrm{max}}$ and thus illustrate the main limitation of the present method.}
    \label{fig:combined_n=0.72_U=8_tprime-0.2}
\end{figure}

In \Cref{fig:combined_n=0.72_U=8_tprime-0.2}(a), we show
$\chi_{\vec{q}}$ along the high-symmetry path and decompose it into its bubble and vertex contributions. The vertex contribution exceeds the bubble by roughly an order of magnitude. Moreover, the two carry a distinct momentum structure (Appendix~\ref{sec:app_doping} contains results at other dopings). We find qualitative agreement, regarding the position and structure of the incommensurate peaks and their evolution with doping, with the results of Ref.~\onlinecite{vilardi2018} obtained with an exact-diagonalization (ED) impurity solver. There are, however, differences in the magnitude of $\chi_{\vec{q}}$. We have confirmed with the authors of Ref.~\onlinecite{vilardi2018} that this quantitative discrepancy probably originates in the discretization of the ED solver employed there.
Newly generated vertex results from a continuous-time hybridization-expansion (CT-HYB) QMC solver \cite{wallerberger2019} [dashed lines in \cref{fig:combined_n=0.72_U=8_tprime-0.2}(a)] agree quantitatively with our results.

Finally, we study significantly lower temperatures,
which are inaccessible to QMC-based solvers but become accessible with iDMFT (removing the need to compute the vertex) using an impurity solver like NRG.
\Cref{fig:combined_n=0.72_U=8_tprime-0.2}(b) shows the lattice susceptibility %of the doped Hubbard model with 
along the high-symmetry path
at the same parameters as before
($U=8$, $n=0.72$, $t'=-0.2$),
from the moderate temperature $\beta=12.5$ studied in Ref.~\onlinecite{vilardi2018} down to an unprecedented $\beta=100$. 
On lowering the temperature, the response sharpens into a pronounced peak; the peak does not, however, shift much with temperature. As an independent check, we again mark the homogeneous single-site DMFT susceptibility $\chi_{\vec{q}=0}$ at the $\Gamma$ point (see also the inset).

Calculations at these low temperatures expose the main limitation of the method. Since iDMFT evaluates $\chi_i$ in a finite system of size $d_{\mathrm{max}}$, the calculation is accurate only if $\chi_i$ has decayed sufficiently by $d_{\mathrm{max}}$. The magnetic correlation length typically grows with lowering the temperature; once it exceeds $d_{\mathrm{max}}$, the Fourier transform truncates a slowly decaying tail, thus underestimating the peak in $\chi_{\vec{q}}$ and creating finite-size artifacts. 
One could remedy this by extrapolating $\chi_i$ to larger $|\vec{r}_i|$, but we refrain from doing so here.

\Cref{fig:combined_n=0.72_U=8_tprime-0.2}(c) shows the real-space decay of $\chi_i$ for $\vec{r}_i=(x, 0)$ and $\beta=50,\,100,\, 500$ on a fairly large system with $d_{\mathrm{max}}=89$, corresponding to 4095 independent impurities at each iDMFT iteration. For $\beta=50,\,100$, it decays to below $10^{-4}$ at $d_{\mathrm{max}}$, while it remains about two orders of magnitude higher at $d_{\mathrm{max}}$ for $\beta=500$. This means that the system size is not large enough for this temperature. The inset in panel (c) quantifies this by comparing $\chi_i$ from iDMFT with system sizes $d_{\mathrm{max}}=70$ (2556 impurities) and $d_{\mathrm{max}}=89$ (4095 impurities). For $\beta=50,\,100$, the results agree to $10^{-4}$, confirming convergence in system size, while the difference for $\beta=500$ is on average two orders of magnitude larger. We therefore omit the $\beta=500$ result from panel (b), but show it in (c) to explain the finite-size limitation.

\section{Conclusions}
\label{sec:conclusions}

We have presented a method for computing static lattice spin susceptibilities in DMFT without requiring two-particle vertex functions. The method is based on inhomogeneous DMFT and generates the susceptibility from the linear response to a small but finite magnetic field, localized in space. An expansion of all iDMFT equations in the external field avoids large matrix inversion, and a resummation scheme recovers the DMFT magnetic instability in the thermodynamic limit. We showed analytically that the results obtained with the DMFT vertex approach and with iDMFT are equivalent.

The iDMFT approach to lattice susceptibilities is highly versatile: It can be combined with virtually any impurity solver and, by using a different external field (e.g., a local variation of the chemical potential instead of the magnetic field), one obtains another static susceptibility (e.g., the density instead of magnetic susceptibility).
Its limitation lies in the cost of large system sizes, needed to resolve large (magnetic) correlation lengths. 
As a cheap alternative, 
we have also introduced a rank-1 approximation of the local vertex, which improves the RPA at negligible extra computational cost. 

We have benchmarked our approach against the standard vertex formalism and demonstrated good agreement.
Using NRG as the impurity solver, the method can treat finite doping at no extra cost and can reach low temperatures that are inaccessible to the standard vertex approach.
Hence, it eliminates the numerical bottleneck of many previous studies \cite{Strand2019,Adler2024}
and holds great promise for studying lattice susceptibilities of multiorbital systems (where vertex calculations at low temperature become prohibitively expensive), both in the model \cite{Stadler2015,Kugler2019,Kugler2022_OSMP} and material contexts \cite{Kugler2020,Kugler2024,Grundner2025,Kugler2026,LaBollita2026}.

\begin{acknowledgments}
We thank Antoine Georges, Dominik Kiese, Harrison LaBollita, Jae-Mo Lihm, Nils Wentzell for useful discussions and Seung-Sup Lee and Rok Žitko for a critical reading of the manuscript.
Furthermore, we thank Ruslan Mushkaev, Philipp Werner, and Demetrio Vilardi for a helpful correspondence and, especially, Demetrio Vilardi for generating the QMC results in Fig.~\ref{fig:combined_n=0.72_U=8_tprime-0.2}(a).
DR acknowledges the support of the Slovenian Research and Innovation Agency (ARIS) under P1-0416. FBK acknowledges funding from the Ministerium f\"ur Kultur und Wissenschaft des Landes Nordrhein-Westfalen (NRW-R\"uckkehrprogramm). 
This work was initiated as part of the Pre-Doctoral Program at the Center for Computational Quantum Physics at the Flatiron Institute. 
The Flatiron Institute is a division of the Simons Foundation. 
\end{acknowledgments}

\section*{Data availability}
The data that support the results of this study are openly available on a Zenodo repository \cite{rolih2026_data}. All software used is freely available \cite{nrgljubljana, parcollet2015, triqs_ctint, triqs_tprf}.

\appendix
\section{Numerical details}
\label{sec:numerical_details}

This appendix summarizes the iDMFT algorithm together with the relevant choices of the parameters. The algorithm can be summarized as follows: First, run a single-site DMFT calculation in the absence of the perturbation $V^{(i)}_{\sigma} = 0$.
With the self-energy $\Sigma(z)$ from single-site DMFT, construct the single-site DMFT lattice Green functions in momentum-space $G_\vec{k}(z)$ [Eq.~\eqref{eq:interacting_lattice_green_k_space}] and real-space 
$G_{i-j}(z)$ [Eq.~\eqref{eq:fourier_transform_G}].
Second, initialize the iDMFT impurity models in the presence of a small, localized external field (here magnetic field) $V^{(i)}_{\sigma} \propto \delta_{i,0} \neq 0$ by Eq.~\eqref{eq:iDMFT_Weiss_field} with $\tilde{\Sigma}_\sigma^{(j)}(z)=0$. After iDMFT convergence, compute the expectation value of the operator conjugate to the external field (here $S^z$) to obtain the susceptibility.
Leveraging the single-site DMFT solution for the initialization/update of iDFMT via Eq.~\eqref{eq:iDMFT_Weiss_field}
is crucial to avoid the inversion of a large matrix, see \cref{eq:diagonal_elements_G_int}. The iDMFT system size must be large enough to ensure a proper decay of all quantities to their equilibrium (i.e., $V^{(i)}_{\sigma} = 0$) values ($\tilde{\Sigma}_\sigma^{(j)}(z)\approx 0$, $\langle S_{j}^z\rangle \approx 0$ for $|j|=d_{\mathrm{max}}$). Importantly, the field strength should be small enough to remain within the linear response regime but large enough so that the induced response can be distinguished from the numerical noise of the impurity solver.

We use the full-density-matrix NRG (fdmNRG)~\cite{weichselbaum2007_sum_rule_conserving_spectral_functions_NRG_PhysRevLett,Peters2006} to solve the impurity problems in iDMFT, as implemented in the NRG Ljubljana code \cite{nrgljubljana}.
We use the discretization scheme from Ref.~\onlinecite{zitko2009_discretization_artefacts_PhysRevB} and the improved self-energy estimators of Ref.~\onlinecite{kugler2022_self_energy_trick_PhysRevB}. We use a discretization parameter $\Lambda=2$ and only one discretization grid $N_z=1$ with log-Gaussian broadening parameter $\alpha=0.6$. We have checked that using more discretization grids with smaller $\alpha$ does not qualitatively change the results. During iterative diagonalization, we exploit $U(1)$ charge and $U(1)$ spin symmetry and keep the states corresponding to energies up to $E = 10$ (in units of rescaled energies) with a hard limit of 5000 states.

For single-site and inhomogeneous DMFT convergence, we monitor the integrated difference of the absolute value squared between the local and the impurity Green function on the real frequency axis. We declare the results converged if this difference is less than $10^{-10}$. For iDMFT, we demand this for all impurity problems. This typically occurs in ten to twenty iterations. We have found it useful to use linear mixing, typically with a mixing parameter of $0.5$, to stabilize the DMFT convergence close to magnetic instabilities.

\begin{figure}
    \includegraphics[width=\columnwidth]{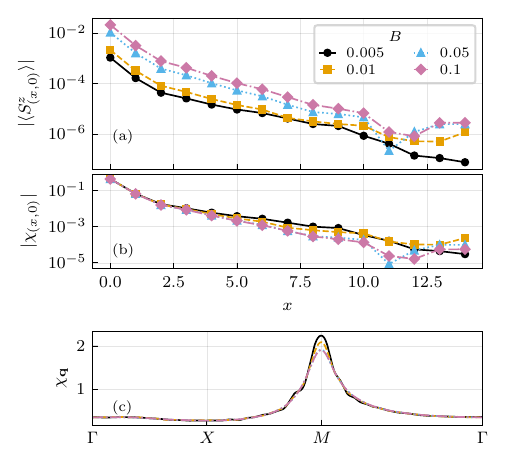}
    \caption{(a) $|\langle S^z_i \rangle|$ and (b) $|\chi_i|$ as a function of $\vec{r}_i=(x,0)$ for various the external-field strengths $B=0.005, 0.05, 0.01, 0.1$, at half-filling, $U=2$, $\beta=5$. (c) Corresponding susceptibility $\chi_{\vec{q}}$. 
    }
    \label{fig:B_sweep}
\end{figure}

The competition between linearity in the external field and signal to noise ratio is examined in \cref{fig:B_sweep}. In panel (a), we show $\langle S^z_i \rangle$ with $\vec{r}_i=(x,0)$ for several field strengths $B$. At small $x$ (distance from the site with applied field), the curves are vertically displaced by exactly the ratio of the applied fields. This is confirmed in panel (b), where we plot $\chi_i$. This means that the use of linear response is justified. 

At larger distances, the exact result would decay exponentially while the numerical $\langle S^z_i\rangle$ reaches the noise threshold (here $\approx 10^{-5}$). This value has to be empirically deduced from the numerical results, and one can tune the parameters of the impurity solver to improve it (see also \cref{fig:maxd_convergence} below). In \cref{fig:B_sweep}(c), we show how the noise enters $\chi_{\vec{q}}$. The curves are indistinguishable near $\vec{q} = 0$ but differ around the AFM peak. The numerical recipe is therefore to choose $B$ as large as possible to enhance the response, while checking that one stays in the linear response regime. In practice, we have used the value of $B=0.05$ in our calculations.

The system size in iDMFT is set by $d_{\mathrm{max}}$, which denotes the furthest lattice site still characterized by a non-vanishing $\tilde\Sigma^{(i)}$. The number of independent impurities to be solved on the square lattice with a perturbation at the origin is $N_{\mathrm{imp}}=(d_{\mathrm{max}}+1)(d_{\mathrm{max}}+2)/2$. We typically choose $d_{\mathrm{max}}= 14$ and thus have $N_{\mathrm{imp}} = 120$. This value was sufficient for most parameters studied here.

In \cref{fig:maxd_convergence}, we show the convergence of the lattice spin susceptibility in system size, characterized by $d_{\mathrm{max}}$. In the present regime (half-filling, $U=2$, $\beta=5$), $d_{\mathrm{max}} = 9$ suffices. Using too small lattice sizes leads to oscillations in $\chi_{\vec{q}}$ and also hinders iDMFT convergence. 
\Cref{fig:maxd_convergence} also confirms that the artifacts in \cref{fig:chi_r_chi_path_U=2}(a) are not a finite-size effect: the results do not change with increasing $d_{\mathrm{max}}$. Rather, they are the consequence of the numerically evaluated $\chi_i$ reaching the noise threshold. This is also the origin of `kinks' for $x > 10$ in \cref{fig:maxd_convergence}(a).

\begin{figure}
    \includegraphics[width=\columnwidth]{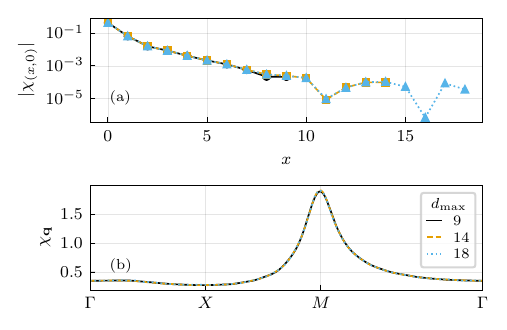}
    \caption{(a) $|\chi_i$ with $\vec{r}_i=(x,0)$ and (b) $\chi_{\vec{q}}$ for different lattice sizes characterized by $d_{\mathrm{max}}$, at half-filling, $U=2$, $\beta=5$.
    }
    \label{fig:maxd_convergence}
\end{figure}

\begin{figure*}
    \centering
    \includegraphics{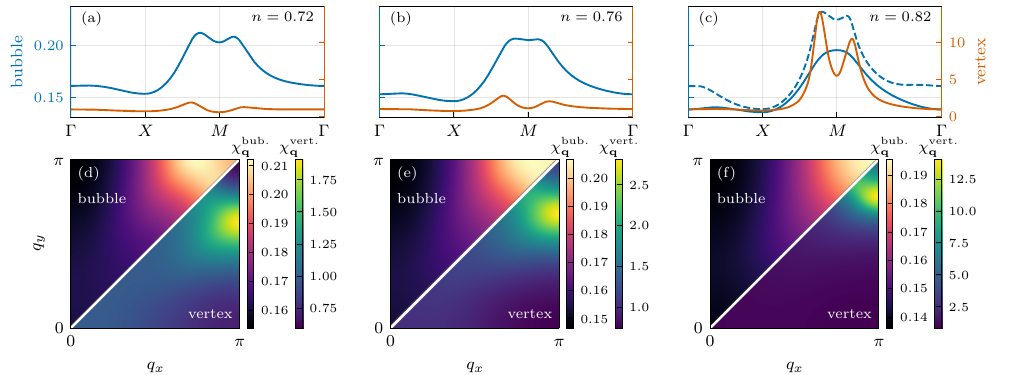}
    \caption{(a)--(c) Bubble and vertex contributions to the lattice spin susceptibility [from \cref{eq:chi_iDMFT_bubble_vertex_momentum_space}] at $U = 8$, $t' = -0.2$, $\beta = 12.5$ for fillings $n = 0.72$, $0.76$, $0.82$. 
    In (c), we also show the bubble contribution at a lower temperature, $\beta=50$ (blue dashed curve).
    (d)--(f) Corresponding plots in the irreducible Brillouin zone (top left triangle: bubble, bottom right triangle: vertex).}
    \label{fig:chi_q_bubble_vertex_combined_U=8_tprime-0.2}
\end{figure*}

\section{Doping dependence at strong coupling}
\label{sec:app_doping}

For completeness, \cref{fig:chi_q_bubble_vertex_combined_U=8_tprime-0.2} shows the doping dependence of the lattice spin susceptibility at $U=8$, $t'=-0.2$, $\beta=12.5$
[\cref{fig:combined_n=0.72_U=8_tprime-0.2}(a) corresponds to \cref{fig:chi_q_bubble_vertex_combined_U=8_tprime-0.2}(a,d)]. This directly reproduces Figures 1 and 2 of Ref.~\onlinecite{vilardi2018}. We reach the same conclusions as they do: at all $n$, the vertex contribution exceeds the bubble by at least an order of magnitude. Furthermore, at $n=0.82$ and the given temperature, the bubble and vertex contributions peak at different values of $\vec{q}$ [see \cref{fig:chi_q_bubble_vertex_combined_U=8_tprime-0.2}(c,f)]. We find that, upon lowering the temperature at $n=0.82$, one recovers the splitting of the $\vec{q}=(\pi, \pi)$ peak also in the bubble.
This is demonstrated by the dashed blue line in \cref{fig:chi_q_bubble_vertex_combined_U=8_tprime-0.2}(c) at $\beta=50$.

\section{Rank-1 approximation of the local vertex}
\label{sec:app_rank1}

\subsection{Formalism}
\label{sec:rank-1}

Equation~\eqref{eq:contracted_WI_1} shows that, for every $\vec{q}$, $\delta\Sigma_{\vec{q}}(\iu\nu)$ and $\chi_{\vec{q}}^{\nu'}$ are related by the $\vec{q}$-independent vertex $\Gamma_{\mathrm{imp}}^{\nu\nu'}$.
Can this relation be used to infer information about the vertex? 
And since the vertex is that of the impurity model, can we use such a strategy already from the single-site DMFT impurity model without invoking the more costly iDMFT (that generates $\delta\Sigma^{(j)}(\iu\nu)$ for many $j$)?
Equation~\eqref{eq:contracted_WI_1} applied to the impurity model reads
\begin{equation}
\label{eq:contracted_WI_approx}
\frac{\delta\Sigma(\iu\nu)}{2B} 
\!=\! 
\frac{1}{\beta}\sum_{\nu'} \Gamma^{\nu\nu'}_{\mathrm{imp}}\chi^{\nu'}_{\mathrm{imp}}
\!\Leftrightarrow\!
\frac{\delta\hat\Sigma(\iu\nu)}{2B}
\!=\!
\frac{1}{\beta}\sum_{\nu'}(\Gamma^{\nu\nu'}_{\mathrm{imp}} + U)\chi^{\nu'}_{\mathrm{imp}}
,
\end{equation}
where we separated the Hartree contribution $\delta\Sigma(\iu\nu) = -2BU\chi_{\mathrm{imp}} + \delta\hat\Sigma(\iu\nu)$ in analogy with \cref{sec:resummation}.
Now, we approximate $\Gamma_{\mathrm{imp}}^{\nu\nu'} + U$ as a rank-1 matrix in frequency space,
\begin{equation}
\label{eq:rank-1_vertex}
\Gamma^{\nu\nu'}_{\mathrm{imp}} \!=\! -U + A^{\nu} \chi^{\nu'}_{\mathrm{imp}}
, \quad
A^{\nu} \!=\! \frac{\delta\hat\Sigma(\iu\nu)}{2B \mathcal{N}}
, \quad
\mathcal{N} \!=\! \frac{1}{\beta}\sum_{\tilde\nu}(\chi^{\tilde\nu}_{\mathrm{imp}})^2
.
\end{equation}
This crude approximation fulfills the high-frequency limit $\Gamma^{\nu\nu'}_{\mathrm{imp}} \to -U$ 
(which would not be the case had we not separated the Hartree part of $\Sigma$)
but violates the $\nu \leftrightarrow \nu'$ symmetry.
Nevertheless, we merely try to approximate $\Gamma^{\nu\nu'}_{\mathrm{imp}}$ in a \emph{weak} sense, i.e., as contracted with $\chi^{\nu}_{\vec{q}}$ [\cref{eq:contracted_WI_1}]. The correct result upon contraction with $\chi^{\nu}_{\mathrm{imp}}$, \cref{eq:contracted_WI_approx}, holds by construction.

One can insert the rank-1 vertex into the lattice BSE, \cref{eq:bethe_salpeter_lattice}, which requires inverting an $N_{\nu} \times N_{\nu}$ matrix at each $\vec{q}$. Due to the rank-1 structure, however, this matrix inversion dramatically simplifies, as we now show.

Inserting \cref{eq:rank-1_vertex} into \cref{eq:bethe_salpeter_lattice_summed}, we get
\begin{equation}
    \label{eq:bse_rank1}
    \chi_{\vec{q}}^{\nu}
    =
    \chi_{0,\vec{q}}^{\nu}
    \left[
   1 + U \chi_{\vec{q}} -  A^{\nu} (\vec{\chi}_{\mathrm{imp}}\cdot\vec{\chi}_{\vec{q}})
    \right],
\end{equation}
where we defined vector quantities $[\vec{\chi}_{\mathrm{imp}}]_{\nu} = \chi_{\mathrm{imp}}^{\nu}$, etc.,
and an inner product $\vec{a}\cdot\vec{b}=\beta^{-1}\sum_{\nu}a^{\nu}b^{\nu}$.
Next, \cref{eq:bse_rank1} implies
\begin{align}
&
\vec{\chi}_{\mathrm{imp}}\cdot\vec{\chi}_{\vec{q}}
=
\vec{\chi}_{\mathrm{imp}}  \cdot \vec{\chi}_{0,\vec{q}} (1 + U \chi_{\vec{q}})
-
(\vec{\chi}_{0,\vec{q}}\cdot\vec{B}) (\vec{\chi}_{\mathrm{imp}}\cdot\vec{\chi}_{\vec{q}})
\nonumber \\
& \Leftrightarrow
\vec{\chi}_{\mathrm{imp}}\cdot\vec{\chi}_{\vec{q}}
=
\frac{
\vec{\chi}_{\mathrm{imp}}  \cdot \vec{\chi}_{0,\vec{q}} (1 + U \chi_{\vec{q}})
}{
1 + \vec{\chi}_{0,\vec{q}}\cdot\vec{B}
}
,
\label{eq:bse_rank1_summed_2}
\end{align}
with $[\vec{B}]_{\nu} \!=\! A^{\nu} \chi_{\mathrm{imp}}^{\nu}$.
Then, \cref{eq:bse_rank1} summed over $\nu$ becomes
\begin{align}
\label{eq:bse_rank1_solved}
\chi_{\vec{q}}
& =
\chi_{0,\vec{q}} ( 1 + U \chi_{\vec{q}} )
-
(\vec{\chi}_{0,\vec{q}}\cdot \vec{A}) (\vec{\chi}_{\mathrm{imp}}\cdot\vec{\chi}_{\vec{q}})
\nonumber \\
& =
( 1 + U \chi_{\vec{q}} )
\bigg[
\underbrace{
\chi_{0,\vec{q}} 
-
\frac{
(\vec{\chi}_{0,\vec{q}}\cdot \vec{A}) 
(\vec{\chi}_{\mathrm{imp}}  \cdot \vec{\chi}_{0,\vec{q}} )
}{
1 + \vec{\chi}_{0,\vec{q}}\cdot\vec{B}
}
}_{
\tilde\chi_{0,\vec{q}} 
}
\bigg]
.
\end{align}
Evidently, this is solved by
\begin{align}
\chi_{\vec{q}} & = \frac{\tilde\chi_{0,\vec{q}}}{1 - U \tilde\chi_{0,\vec{q}}}.
\end{align}
The rank-1 susceptibility takes an RPA form, with the bubble $\chi_{0,\vec{q}}$ replaced by a partially \enquote{vertex-corrected} bubble $\tilde\chi_{0,\vec{q}}$.

In this approach,
solving for the lattice susceptibility $\chi_{\vec{q}}$ at every $\vec{q}$ merely involves constructing the bubble $\chi_{0,\vec{q}}^{\nu}$ (one sum over $\vec{k}$ for each $\nu$) and the Matsubara summations in $\tilde\chi_{0,\vec{q}}$. 
The rank-1 ansatz thus improves on the RPA approach with $\Gamma_{\mathrm{imp}} = -U$ at the cost of a one extra single-site impurity calculation with finite $B$ field, needed to get $\chi^{\nu}_{\mathrm{imp}} = [G_{\uparrow}(\iu\nu)-G_{\downarrow}(\iu\nu)]/(2B)$ and $\delta\hat{\Sigma}(\iu\nu)$ (and thereby $A^\nu$ and $B^\nu$).

\subsection{Numerical results}
\label{sec:res_rank1}

\begin{figure}
    \centering
    \includegraphics{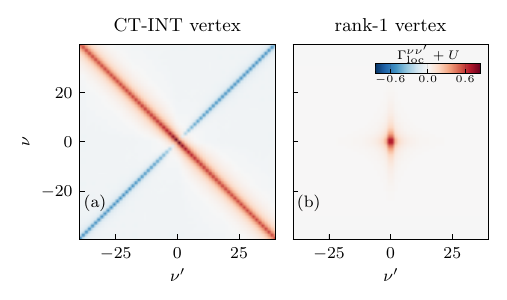}
    \caption{(a) Exact vertex $\Gamma_{\mathrm{imp}}^{\nu\nu'}+U$ and (b) rank-1 approximation at $U=2$, $\beta=5$ and half-filling. The vertex in (a) is obtained by inverting the two-particle Green function (obtained with a CT-INT solver) in the impurity BSE. The vertex in (b) is constructed from spin-dependent self-energy and Green function obtained by one additional solution of the impurity model with finite magnetic field (solved with NRG). Both panels have the same color scale and use $32$ fermionic Matsubara frequencies $\nu,\nu'$.}
    \label{fig:vertex_exact_vs_rank1}
\end{figure}

First, \cref{fig:vertex_exact_vs_rank1} shows the frequency dependence of the impurity vertex obtained by CT-INT and its rank-1 approximation at $U=2$, $\beta=5$, $n=1$. The rank-1 approximation misses the characteristic diagonal/anti-diagonal structure of vertex. However, as noted before, to compute $\chi_{\vec{q}}$, we merely want to approximate the vertex as contracted with $\chi_{0,\vec{q}}$.

\begin{figure}[ht]
    \centering
    \includegraphics{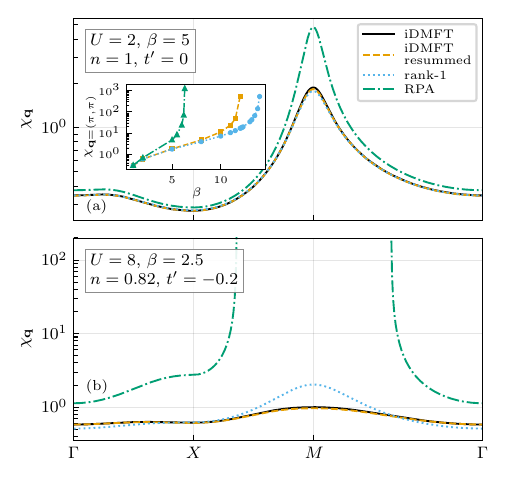}
    \caption{Lattice spin susceptibility for (a) $U=2$, $\beta=5$, $n=1$, $t'=0$ and (b)  $U=8$, $\beta=2.5$, $n=0.82$, $t'=-0.2$. Inset in panel (a): divergence of $\chi_{\vec{q}}$ at the magnetic ordering vector $\vec{q}=(\pi, \pi)$ with inverse temperature $\beta$. The parameters in panel (b) are chosen such that RPA already reaches an instability, while the rank-1 approximation gives a well-behaved $\chi_{\vec{q}}$, albeit overestimating $\chi_{\vec{q}=M}$. The legend is shared between all panels.}
    \label{fig:rank1_compare_U}
\end{figure}

We test the results of the rank-1 approximation for $\chi_{\vec{q}}$ in two different parameter regimes: at weak interaction ($U=2$, $\beta=5$, $n=1$, $t'=0$) and at strong interaction ($U=8$, $\beta=2.5$, $n=0.82$, $t'=-0.2$). The latter parameter set was chosen so that the rank-1 approximation produces a physical $\chi_{\vec{q}}$ (i.e., $0<\chi_{\vec{q}}<\infty$), which, however, differs notably from the full iDMFT.
The results for $\chi_{\vec{q}}$ in both regimes are shown in
\cref{fig:rank1_compare_U}(a,b). They are obtained with iDMFT, iDMFT resummation [\cref{eq:chi_q_psi_corrected}], the rank-1 approximation [\cref{eq:rank-1_vertex}], and RPA. In the weak-interaction regime of (a), the rank-1 approximation works remarkably well and is virtually equivalent to the full iDMFT calculation. We have established in \cref{sec:res_weak_coupling} that iDMFT gives the same result as the calculation of $\chi_{\vec{q}}$ through the BSE with the exact impurity vertex. We here see that the contracted vertex is well reproduced by its crude rank-1 approximation. Still, some frequency dependence of the vertex must be included, as the RPA result, originating from $\Gamma_{\mathrm{imp}} = -U$, strongly overestimates $\chi_{\vec{q}}$.

In the strong-interaction regime of \cref{fig:rank1_compare_U}(b), 
the RPA already gives an unphysical, negative $\chi_{\vec{q}}$ for a range of $\vec{q}$ around $(\pi,\pi)$. In contrast, the rank-1 approximation results in a physical, positive $\chi_{\vec{q}}$ but strongly overestimates its value. This shows that the rank-1 approximation has a limited range of validity; it is prone to overestimating the magnetic response but much less so than RPA.
We expect the rank-1 approximation to work well if all $\chi^\nu_{0,\vec{q}}$ are similar to $\chi^\nu_{\mathrm{imp}}$ (it should work perfectly if they are linearly dependent). 
Away from half-filling, $\chi^\nu_{0,\vec{q}}$ and $\chi^\nu_{\mathrm{imp}}$ become complex-valued, so that such a similarity (or, ideally, linear dependence) is less likely.

\Cref{fig:rank1_compare_U} thus summarizes the main strengths and weaknesses of the rank-1 approximation:
It can be viewed as an inexpensive correction over the RPA (setting $A^{\nu}=0$ in \cref{eq:rank-1_vertex} recovers the RPA static vertex $\Gamma_{\mathrm{imp}}=-U$), improving the quantitative agreement of the lattice susceptibility $\chi_{\vec{q}}$ at high temperatures and shifting the magnetic instability to lower temperatures [see inset to \cref{fig:rank1_compare_U}(a)]. Nevertheless, it cannot fully replace the DMFT vertex or a proper iDMFT calculation: apparently, the full frequency structure of the vertex or the site-dependent response in iDMFT are needed for accurate results at strong interaction and/or finite doping [see \cref{fig:rank1_compare_U}(b)].

\section{2D square lattice DOS}
\label{sec:2d_dos}

We derive the density of states (DOS) of the square lattice with nearest-neighbor hopping $t$ and next-nearest neighbor hopping $t'$. For $t'=0$, the result is well-known; for $t' \neq 0$, we found \cref{eq:2d_dos} in the literature~\cite{zhuravlev2011}. Yet, this formula holds only if $|t'| < t/2$. Here, we generalize it to arbitrary $t'$.

The definition of the DOS is $\rho(\epsilon) = \frac{1}{(2\pi)^2}\int_{-\pi}^{\pi}\dif k_x \int_{-\pi}^{\pi} \dif k_y \delta(\epsilon - \varepsilon_{\vec{k}})$ with $\varepsilon_{\vec{k}}$ given in \cref{eq:2d_dispersion}. We start by integrating out $k_y$; define $A(u) = 2t + 4t'u$ and $B(u) = \epsilon + 2tu$ with $u=\cos k_x$, in anticipation of a variable change later. The $\delta$-function sets $\cos k_y = -B/A$. On $[-\pi, \pi]$, there are two such solutions, $\pm k_y$, both satisfying
%\begin{subequations}
\begin{align}
    %&|\sin k_y| = \sqrt{1 - \cos^2 k_y}, \\
    &\Big| \dod{\epsilon_{\vec{k}}}{k_y} \Big| = |2t + 4t'\cos k_x| |\sin k_y| = |A|\sqrt{1 - \cos^2 k_y}.
\end{align}
%\end{subequations}
Both solutions give equal contributions, and we get
\begin{align}
    \int_{-\pi}^{\pi} \dif k_y \delta(\epsilon - \varepsilon_{\vec{k}})
    &=
    \int_{-\pi}^{\pi}\dif k_y \sum_{i} \frac{\delta(k_y - k_y^{(i)})}{|\dif\epsilon/\dif k_y|}
    \nonumber \\
    &=
    \frac{2}{|A|\sqrt{1-\cos^2 k_y}}.
\end{align}

Now, we substitute $u=\cos k_x$, $k_x \in [-\pi, \pi]$, with $\dif k_x = -\dif u / \sqrt{1 - u^2}$. There are two values of $k_x$ for each $u\in(-1, 1)$, and we get
\begin{equation}
    \int_{-\pi}^{\pi} \dif k_x f(\cos k_x) = 2 \int_{-1}^{1} \frac{\dif u}{\sqrt{1-u^2}} f(u).
\end{equation}
This leaves us with a one-dimensional integral
\begin{equation}
    \rho(\epsilon) = \frac{1}{\pi^2}\int_{-1}^{1} \frac{\dif u}{\sqrt{(1-u^2)(A^2 - B^2)}}.
\end{equation}

The term under the square root has four roots: $\pm 1$ and
\begin{equation}
    u_{-} = \frac{\epsilon-2t}{2(2t' - t)},\; u_{+} = - \frac{2t + \epsilon}{2(2t' + t)}.
\end{equation}
This quartic polynomial can thus be written as
\begin{equation}
    Q(u) = -\alpha (u + 1)(u - 1)(u - u_{+})(u - u_{-})
\end{equation}
with the leading coefficient $\alpha = 4 (4t'^2 - t^2)$, so that
\begin{equation}
    \rho(\epsilon) = \frac{1}{\pi^2}\int_{-1}^{1} \frac{\dif u}{\sqrt{Q(u)}}.
\end{equation}
We need to find those subintervals of $[-1, 1]$ on which $Q(u) > 0$ since only those will contribute to the DOS.

Let us sort the four roots as $r_1 \!<\! r_2 \!<\! r_3 \!<\! r_4$. First, since $\pm1$ are always among the roots, we have $r_1 \leq -1$ and $r_4 \geq 1$. This means that the intervals $(-\infty, r_1)$ and $(r_4, \infty)$ lie outside the integration region $[-1, 1]$. Second, since $Q$ is continuous, its sign is constant between consecutive roots and alternates.

The sign of $Q$ on each of the contributing subintervals is
\begin{equation}
    \label{eq:sign_Q}
    \sign Q = 
    \begin{cases}
    \sign \alpha  &u \in (r_1, r_2) \cup (r_3, r_4),\\
    -\sign \alpha &u \in (r_2, r_3).
    \end{cases}
\end{equation}
In each case where $Q(u) > 0$, we can write the polynomial as $Q(u) = |\alpha|p(u)$, where $p$ is a product of positive factors. Explicitly, on $(r_1, r_2)$ with $\alpha>0$
\begin{equation}
    Q(u) = |\alpha|(u-r_1)(r_2-u)(r_3-u)(r_4-u);
\end{equation}
on $(r_2, r_3)$ with $\alpha < 0$
\begin{equation}
    Q(u) = |\alpha|(u-r_1)(u-r_2)(r_3-u)(r_4-u);
\end{equation}
and on $(r_3, r_4)$ with $\alpha > 0$
\begin{equation}
    Q(u) = |\alpha| (u-r_1)(u-r_2)(u-r_3)(r_4-u).
\end{equation}
The integral on any contributing interval can thus be written in the form
\begin{equation}
    \int_{r_i}^{r_{i+1}} \frac{\dif u}{\sqrt{Q(u)}} = \frac{1}{\sqrt{|\alpha|}}\int_{r_i}^{r_{i+1}} \frac{\mathrm{d}u}{\sqrt{p(u)}}.
\end{equation}

For $|t'| < t/2$ we have $\alpha < 0$; this means that according to \cref{eq:sign_Q} only the integral over the interval $(r_2, r_3)$ contributes to the DOS.
To see that this reduces to \cref{eq:2d_dos}, we evaluate the prefactor and the function argument in terms of $t$, $t'$, and $\epsilon$. 
First, compute
\begin{equation}
    u_{-} - u_{+} = \frac{8(\epsilon t' - t^2)}{\alpha}.
\end{equation}
Since $\alpha < 0$ and $t^2 - \epsilon t' > 0$ inside the band, this is positive, and thus $u_{-} > u_{+}$. For $|t'| < t/2$, at any energy
in the band exactly one of $u_\pm$ lies inside $(-1,1)$ and the
other outside. In both cases, the roots satisfy
$(r_3-r_1)(r_4-r_2) = 2(u_- - u_+)$, giving
\begin{equation}
    |\alpha|(r_3-r_1)(r_4-r_2) = 16(t^2 - \epsilon t').
\end{equation}
Secondly, the function argument evaluates to
\begin{equation}
    m_{23}
    = \frac{(1-u_+)(1+u_-)}{2(u_- - u_+)}
    = \frac{t^2 - (\epsilon + 4t')^2/16}{t^2 - \epsilon t'}.
\end{equation}
Together, this reproduces \cref{eq:2d_dos}:
\begin{equation}
    \rho(\epsilon)
    = \frac{1}{2\pi^2\sqrt{t^2 - \epsilon t'}}\,
      K\left(\sqrt{\frac{t^2 - (\epsilon+4t')^2/16}{t^2 - \epsilon t'}}\right).
\end{equation}

For $|t'| > t/2$, we need to evaluate the integral over the other two intervals. We show how this can be done for $(r_1, r_2)$ (the other cases are analogous). The integral is
\begin{equation}
    I_{12} = \int_{r_1}^{r_2} \frac{\dif u}{\sqrt{(u-r_1)(r_2 - u)(r_3-u)(r_4-u)}}
\end{equation}
and can be transformed to an elliptic integral. Indeed, substituting $u = r_1 + (r_2 - r_1) \sin^2\varphi$ and simplifying gives
\begin{equation}
    I_{12} = \frac{2}{\sqrt{(r_3-r_1)(r_4-r_1)}} J(a, b)
\end{equation}
with 
\begin{equation}
    J(a, b) = \int_{0}^{\pi/2}\frac{\dif \varphi}{\sqrt{(1 - a\sin^2\varphi)(1 - b \sin^2 \varphi)}}
\end{equation}
and
\begin{equation}
    a = \frac{r_2 - r_1}{r_3 - r_1},\; b = \frac{r_2 - r_1}{r_4 - r_1}
.
\end{equation}

Changing variables once more according to
\begin{equation}
    \sin^2\varphi = \frac{\sin^2\psi}{1 - a \cos^2 \psi}
,
\end{equation}
we get with $m=(a - b) / (1 - b)$
\begin{equation}
    J(a, b) = \frac{1}{\sqrt{1 - b}}\int_{0}^{\pi/2}\frac{\dif \psi}{\sqrt{1 - m \cos^2 \psi}}.
\end{equation}
Finally, shifting the angle $\chi = \pi/2 - \psi$ gives us
\begin{equation}
    J(a, b) = \frac{1}{\sqrt{1 - b}}\int_{0}^{\pi/2} \frac{\dif \chi}{\sqrt{1 - m \sin^2 \chi}} = \frac{K(\sqrt{m})}{\sqrt{1-b}}
\end{equation}
where $K$ is the complete elliptic integral of the first kind, defined as $K(k)=\int_{0}^{\pi/2}\dif\vartheta/\sqrt{1 - k^2 \sin^2\vartheta}$.

The final result, expressed with the roots $r_i$, is
\begin{subequations}
\begin{align}
    &I_{12} = \frac{2}{\sqrt{(r_3 - r_1)(r_4 - r_2)}}K(\sqrt{m_{12}}),\\
    &m_{12} = \frac{(r_2 - r_1)(r_4 - r_3)}{(r_3 - r_1)(r_4 - r_2)}.
\end{align}
\end{subequations}
An analogous derivation for the interval $(r_2, r_3)$ leads to
\begin{subequations}
\begin{align}
    &I_{23} = \frac{2}{\sqrt{(r_3 - r_1)(r_4 - r_2)}} K(\sqrt{m_{23}}),\\
    &m_{23} = \frac{(r_3 - r_2)(r_4 - r_1)}{(r_3 - r_1)(r_4 - r_2)} = 1 - m_{12};
\end{align}
\end{subequations}
and for $(r_3, r_4)$, using $u=r_3 + (r_4 - r_3)\sin^2\varphi$ produces $\cos^2\varphi$ factors instead of $\sin^2\varphi$; after $\varphi \to \pi/2 - \varphi$, the derivation is identical and gives $I_{34} = I_{12}$.
Each contributing interval therefore gives
\begin{equation}
    \rho_{i}(\epsilon) = \frac{1}{\pi^2 \sqrt{|\alpha|}} I_i
\end{equation}
and the total DOS is the sum of all these contributions.

\bibliography{references}

\end{document}